\documentclass[12pt]{article}
\usepackage{a4wide}
\usepackage{epsfig}
\usepackage{amsmath}

\newlength{\absize}
\catcode`@=11
\def\citer{\@ifnextchar [{\@tempswatrue\@citexr}{\@tempswafalse\@citexr[]}}

\def\@citexr[#1]#2{\if@filesw\immediate
  \write\@auxout{\string\citation{#2}}\fi
  \def\@citea{}\@cite{\@for\@citeb:=#2\do
    {\@citea\def\@citea{--\penalty\@m}\@ifundefined
       {b@\@citeb}{{\bf ?}\@warning
       {Citation `\@citeb' on page \thepage \space undefined}}%
\hbox{\csname b@\@citeb\endcsname}}}{#1}}
\catcode`@=12

\begin{document}
  \thispagestyle{empty}
  \pagestyle{empty}
  \renewcommand{\thefootnote}{\fnsymbol{footnote}}
\newpage\normalsize
    \pagestyle{plain}
    \setlength{\baselineskip}{4ex}\par
    \setcounter{footnote}{0}
    \renewcommand{\thefootnote}{\arabic{footnote}}
\newcommand{\preprint}[1]{%
  \begin{flushright}
    \setlength{\baselineskip}{3ex} #1
  \end{flushright}}
\renewcommand{\title}[1]{%
  \begin{center}
    \LARGE #1
  \end{center}\par}
\renewcommand{\author}[1]{%
  \vspace{2ex}
  {\Large
   \begin{center}
     \setlength{\baselineskip}{3ex} #1 \par
   \end{center}}}
\renewcommand{\thanks}[1]{\footnote{#1}}
%\begin{flushright}
%Oct. 2008
%\end{flushright}
\vskip 0.5cm

\begin{center}
{\large \bf Noiseless Quantum Transmission of Information via
Aharonov - Bohm Effect}
%{\large \bf Non-Dampening and Noiseless
%Quantum Teleportation via Aharonov - Bohm effect}
\end{center}
\vspace{1cm}
%%%%%%
\begin{center}
Jian-Zu Zhang
\end{center}
%-----------------------------------
%   Address
%-----------------------------------
%\vspace{1cm}
\begin{center}
Institute for Theoretical Physics, East China University of
Science and Technology, Box 316, Shanghai 200237, P. R. China
\end{center}
\vspace{1cm}
%%%%%%%%%%%%%%%%%%%%%%%%%%%%%%%%%%%%%%%%%%%%%%%%%%%%%%%%%%%%%%%

\begin{abstract}
The possibility of quantum transmission of information via the
induced fractional angular momentum by the Aharonov - Bohm vector
potential is revealed.
Its special advantage is that it is noiseless:
Stray magnetic
fields of environments influence the energy spectrum of the ion,
but cannot contribute the fractional angular momentum to cause
noise.
%
%There is no energy transmission via the AB vector potential, thus
%no interchange of energy with environments.
%%
%Transmission via an AB vector potential cannot be dampened.
\end{abstract}

%\begin{flushleft}
%$^{\ast}$
%\end{flushleft}
%\clearpage
%%%%%%%%%%%%%%%%%%%%%%%%%%%%%%%%%%%%%%%%%%%%%%%%%%%%%%%%%%%%%%%%%%
Quantum teleportation \citer{BBCJPW,BBMHP} allows the quantum
state of a system to be transported from one location to another,
without moving through the intervening space.
One of its main problems is how to reduce noise which originates
from couplings of the system with uncontrollable environments,
though several ways to reduce noise have been designed.
% including
%quantum error correction,
%%where quantum information is protected by using
%%a certain encoding,
%quantum error avoiding schemes as well as
%entanglement purification.
%
In this paper by means of the Aharonov-Bohm (AB) effect
\cite{ES,AB} we show the possibility of quantum transmission of
information via
%the induced fractional angular momentum by
this effect.
%%%%%%%%%%%%%%%%%%%%%%%%%%%%%%%%%%%%%%%%%%%%%%%%%%%%%%%%%%%%%%%%%%%

The AB effect is purely quantum mechanical phenomenon which has
been received much attention for years \citer{OP85,M-K04}.
%\citer{OP85,WY75}.
%
Experiments \citer{Cham,TOMKEYY} showed that the interference
spectrum of charged particles in a multiply connected region of
the space, where the field strength is zero everywhere, suffered a
shift according to the quantum phase, i.e. the amount of the loop
integral of the magnetic vector potential around an unshrinkable
loop.
It is noticed that the AB effects is due to the non-trivial
topology of a multiply connected region of the space where the
magnetic field strength is vanishing \citer{WY75}.
There are lots of works concerning the fractional angular
momentum:
Investigated from crossed electric and magnetic fields
\cite{PT89,LP82};
Concerned in AB dynamics and their fractional statistics (see the
reviews \citer{Wilc,Laug} and references therein);
Originated from Spatial noncommutativity \cite{JZZ04};
and in connection with the quantum optics \cite{Kast}.
Recently a ``spectator" mechanism \citer{JZZ08} of an induced
fractional angular momentum on ions of the AB vector potential was
revealed.
The ``spectator" mechanism shows that when there is a ``spectator"
magnetic field the AB vector potential in the well defined limit
induces a fractional angular momentum at the full quantum
mechanical level.
This opens a new way of the quantum transmission of information
which is investigated in this paper.
The special advantage of this type of the transmission via the AB
effect is that it is noiseless.
Stray magnetic fields of environments influence the energy
spectrum of the ion, but cannot change the fractional angular
momentum to cause noise.
This effect explores far-reaching consequences of the vector
potential in quantum theory:
The vector potential itself has physical significant meaning and
becomes effectively measurable not only in shifts of interference
spectra originated from quantum phases but also in physical
observables.
%%%%%%%%%%%%%%%%%%%%%%%%%%%%%%%%%%%%%%%%%%%%%%%%%%%%%%%%%%%%%%%%%%

We consider three regions in the $(x_1,x_2)$ plane. (1) A circle I
of radius $a_0$ is centered at the origin $0$ of the coordinates.
A homogeneous magnetic field ${\bf B}^{(0)}$ along the $z$-axis is
concentrated inside the circle I: Inside the circle
$(\rho_0\equiv(x_1^2+x_2^2)^{1/2}<a_0)$
$B^{(0)}_{in,z}=B_0$,
$(\rho_0\ge a_0)$ ${\bf B}^{(0)}_{out}=0$.
The corresponding vector potential ${\bf A}^{(0)}$ is (Henceforth
the summation convention is used): Inside the circle
$(\rho_0<a_0)\;$
$A^{(0)}_{in,i}=-B_0\epsilon_{ij}x_j/2$,\; $A^{(0)}_{in,z}=0;$
Outside the circle
$(\rho_0\ge a_0)\;$
\begin{equation}
\label{Eq:A}%1e
A^{(0)}_{out,i}=-B_0 a^2_0\frac{\epsilon_{ij}x_j}{2\rho_0^2}\;
(i,j,k=1,2),\;A^{(0)}_{out,z}=0.
%
%A^{(0)}_{out,i}=-B_0 a^2_0 \epsilon_{ij}x_j/2\rho_0^2\;
%(i,j,k=1,2),\;A^{(0)}_{out,z}=0.
\end{equation}
${\bf A}^{(0)}_{out}$ is a vector potential of the AB type. At
$\rho_0=a_0$ the potential ${\bf A}^{(0)}_{out}$ passes
continuously over into ${\bf A}^{(0)}_{in}$.
(2) A circle II of a radius $a_c$ is  centered at the point $C$ of
coordinators ${\bf x}_C=(x_C,0,0)$. Here $x_C$, $a_0$ and $a_c$
satisfy $x_C>a_0+a_c$.
Inside the circle II
$(\rho_c\equiv[(x_1-x_C)^2+x_2^2]^{1/2}<a_c)\;$ there is a
homogeneous magnetic field
$B^{(c)}_{in,z}=B_c$
along the $z$ - axis, and outside the circle II $(\rho_c\ge a_c)$
${\bf B}^{(c)}_{out}=0$.
The corresponding vector potential ${\bf A}^{(c)}$ is: Inside the
circle II $(\rho_c<a_c)\;$
$A^{(c)}_{in,i}=-B_c\epsilon_{ij}(x_j-x_{C,j})/2$,
$A^{(0)}_{in,z}=0;$
and outside the circle II $(\rho_c\ge a_c)$
$A^{(c)}_{out,i}=-B_c a_c^2 \epsilon_{ij}(x_j-x_{C,j})/2\rho_c^2\;
(i,j,k=1,2),\;
A^{(c)}_{out,z}=0.$
At the circle II $(\rho_c=a_c)\;$ the potential ${\bf
A}^{(c)}_{out}$ passes continuously over into ${\bf
A}^{(c)}_{in}$.
(3) An intervening region III is an area outside the circle I and
II where the coordinates $(x_1,x_2)$ of a point $P$ satisfy both
conditions
$\rho_0\ge a_0$
and
$\rho_c\ge a_c$.
In the region III the magnetic fields
${\bf B}^{(0)}={\bf B}^{(c)}=0,$
but there are two vector potentials of the AB type:
${\bf A}^{(0)}_{out}$
and
${\bf A}^{(c)}_{out}$.

{\bf Induced Fractional Angular Momentum by the AB Vector
Potential} -- We consider an ion with mass $\mu$ and charge
$q(>0)$ constrained in the circle II where the vector potentials
are $A^{(c)}_{in,i}$ and $A^{(0)}_{out,i}$.
In the following in $A^{(c)}_{in,i}$ we don't consider the
constant term $B_c\epsilon_{ij}x_{C,j}/2$ which can be gauged away
by a gauge transformation
$A^{(c)}_{in,i}\to A^{(c)}_{in,i}+\partial_i
\chi=-B_c\epsilon_{ij}x_j/2$ with $\chi=-B_c\epsilon_{ij}x_i
x_{C,j}/2$.
The Hamiltonian of the charge particle is
\begin{equation}
\label{Eq:H1}%2e
H(x_1,x_2)=\frac{1}{2\mu}\left(p_i+\frac{1}{2}\mu\omega_c
\epsilon_{ij}x_j+\mu\omega_0
a^2_0\frac{\epsilon_{ij}x_j}{2\rho_0^2}\right)^2,
\end{equation}
where $\omega_c=qB_c/\mu c$ and $\omega_0=qB_0/\mu c$ are the
cyclotron frequencies corresponding to, respectively, the magnetic
fields ${\bf B}^{(c)}_{in}$ and ${\bf B}^{(0)}_{in}$. This
Hamiltonian can be rewritten as
$H(x_1,x_2)=\left(K_1^2+K_2^2\right)/2\mu$
where
\begin{equation}
\label{Eq:K1}%3e
K_i\equiv p_i+\frac{1}{2}\mu\omega_c \epsilon_{ij}x_j +\mu\omega_0
a^2_0\frac{\epsilon_{ij}x_j}{2\rho_0^2
},
%
%K_i\equiv p_i+\mu\omega_c \epsilon_{ij}x_j/2 +\mu\omega_0
%a^2_0\epsilon_{ij}x_j/2\rho_0^2,
%\quad
%
%[K_i,K_j]=i\hbar\mu\omega_c\delta_{ij}.
\end{equation}
is the mechanical momenta corresponding to the vector potentials
$A^{(c)}_{in,i}$ and $A^{(0)}_{out,i}$.
The commutation relations between $K_i$'s are
\begin{equation}
\label{Eq:K1}%4e
[K_i,K_j]=i\hbar\mu\omega_c\epsilon_{ij}.
\end{equation}
One point that should be emphasized is that the AB vector
potential $A^{(0)}_{out,i}$ does {\it not} contributes to the
commutator $[K_i,K_j]$. In the above $p_i=-i\hbar\partial/\partial
x_i$ are the canonical momenta, which satisfy $[p_i,p_j]=0$. They
are different from the mechanical momenta $K_i$.
We define canonical variables $Q=K_1/\mu\omega_c$ and $\Pi=K_2$.
They satisfy
$[Q,\Pi]=i\hbar\delta_{ij}.$
The Hamiltonian $H(x_1,x_2)$ is rewritten as one of a harmonic
oscillator,
$H(x_1,x_2)=H(Q,\Pi)=\Pi^2/2\mu+\mu\omega_c^2 Q^2/2.$
Its eigenvalues are
$\mathcal{E}_{n}=\hbar\omega_c(n+1/2)$.
The lowest one is
$\mathcal{E}_0=\hbar\omega_c/2.$
From the lowest eigenvalue $\mathcal{E}_0$  we estimate that the
size $a_c$ of the circle II should satisfy
$a_c\ge \left(c\hbar/qB_c\right)^{1/2}.$

It is worth noting that $A^{(0)}_{out,i}$ does {\it not}
contribute to energy spectra.

The limiting case of the Hamiltonian $H$ in Eq.~(\ref{Eq:H1})
approaching its lowest eigenvalue is interesting. In this limit
the system has non-trivial dynamics \cite{Baxt,JZZ96}. The
Lagrangian corresponding to $H$ is
\begin{equation}
\label{Eq:L1}%5e
L=\frac{1}{2}\mu\dot{x_i}\dot{x_i} -\frac{1}{2}\mu\omega_c
\epsilon_{ij}\dot{x_i}x_j-\mu\omega_0
a^2_0\frac{\epsilon_{ij}\dot{x_i}x_j}{2\rho_0^2}.
\end{equation}
In this limit the Hamiltonian $H$ reduces to
%
%\begin{equation}
%\label{Eq:H}%e
%H_0=\frac{1}{2}\hbar\omega_c + \frac{1}{2}\mu\omega_P^2 x_i
%x_i.
%%%%%%
$H_0=\hbar\omega_c/2.$
%\end{equation}
%
The Lagrangian corresponds to $H_0$ is
\begin{equation}
\label{Eq:L2}%6e
L_0=-\frac{1}{2}\mu\omega_c \epsilon_{ij}\dot{x_i}x_j-\mu\omega_0
a^2_0\frac{\epsilon_{ij}\dot{x_i}x_j}{2\rho_0^2}
-\frac{1}{2}\hbar\omega_c.
\end{equation}

{\it Constraints} -- For the reduced system $(H_0,L_0)$ the
canonical momenta are
\begin{equation}
\label{Eq:p1}%7e
p_{i}=\frac{\partial L_0}{\partial
\dot{x_i}}=-\frac{1}{2}\mu\omega_c \epsilon_{ij}x_j-\mu\omega_0
a^2_0\frac{\epsilon_{ij}x_j}{2\rho_0^2}.
\end{equation}
Eq.~(\ref{Eq:p1}) does not determine velocities $\dot{x_i}$ as
functions of $p_{i}$ and $x_j$, but gives relations among
$p_{i}$'s and $x_j$'s.
According to Dirac's formalism of quantizing constrained system,
such relations are the primary constraints \cite{JZZ96,M-K06}
\begin{equation}
\label{Eq:C1}%8e
\varphi_i\equiv p_i +\frac{1}{2}\mu\omega_c
\epsilon_{ij}x_j+\mu\omega_0
a^2_0\frac{\epsilon_{ij}x_j}{2\rho_0^2}=0.
\end{equation}
These constraints should be carefully treated. The subject can be
treated simply by the symplectic method in \cite{FJ,DJ93}. In this
paper we work in the formalism of the Dirac brackets. The Poisson
brackets of the constraints (\ref{Eq:C1}) are
\begin{equation}
\label{Eq:Poisson-1}%9e
C_{ij}=\{\varphi_i, \varphi_j\}= \mu\omega_c\epsilon_{ij}.
\end{equation}
From Eq.~(\ref{Eq:Poisson-1}), $\{\varphi_i, \varphi_j\}\ne 0,$ it
follows that the conditions of the constraints $\varphi_i$ holding
at all times do not lead to secondary constraints.

$C_{ij}$ defined in Eq.~(\ref{Eq:Poisson-1}) are elements of the
constraint matrix $\mathcal{C}.$ Elements of its inverse matrix
$\mathcal{C}^{-1}$ are $(C^{-1})_{ij}=-\epsilon_{ij}/\mu\omega_c.$
The corresponding Dirac brackets of $\{\varphi_i, x_j\}_D$,
$\{\varphi_i, p_j\}_D$, $\{x_i, x_j\}_D$, $\{p_i, p_j\}_D$ and
$\{x_i, p_j\}_D$ can be defined. The Dirac brackets of $\varphi_i$
with any variables $x_i$ and $p_j$ are zero so that the
constraints (\ref{Eq:C1}) are strong conditions. It can be used to
eliminate dependent variables. If we select $x_1$ and $x_2$ as the
independent variables, from the constraints (\ref{Eq:C1}) the
variables $p_1$ and $p_2$ can be represented by, respectively, the
independent variables $x_2$ and $x_1$ as
\begin{equation}
\label{Eq:p-x-1}%10e
p_1=-\frac{1}{2}\mu\omega_c x_2-\mu\omega_0
a^2_0\frac{x_2}{2\rho_0^2}, \;
p_2=\frac{1}{2}\mu\omega_c x_1+\mu\omega_0
a^2_0\frac{x_1}{2\rho_0^2}
\end{equation}
The Dirac brackets of $x_1$ and $x_2$ is
\begin{equation}
\label{Eq:Dirac}%11e
\{x_1,x_2\}_D=\frac{1}{\mu\omega_c}.
\end{equation}
We introduce new canonical variables $q=x_1$  and $p=\mu\omega_c
x_2.$ Their Dirac bracket is $\{q,p\}_D=1.$ According to Dirac's
formalism of quantizing a system which is associated with a number
of primary constraints, the corresponding quantum commutation
relation is $[q,p]=i\hbar$.

{\it Angular Momentum of the Reduced System} -- The Hamiltonian
$H$ in Eq.~(\ref{Eq:H1}) possess a rotational symmetry in $(x_1,
x_2)$ plane. The $z$-component of the orbital angular momentum
$J_z=\epsilon_{ij} x_i p_j$
commutes with $H$. They have common eigenstates. Now we consider
the quantum behavior of the angular momentum in the reduced system
$(H_0,L_0)$. Using Eq.~(\ref{Eq:p-x-1}) to replace $p_1$ and $p_2$
by, respectively, the independent variables $x_2$ and $x_1$, then
using new canonical variables $p$ and $q$ to replace $x_2$ and
$x_1$, the orbital angular momentum $J_z$ is rewritten as
\begin{equation}
\label{Eq:J}%12e
J_z=\frac{q}{2\pi c}\Phi_0 +
\frac{1}{\omega_c}\left(\frac{1}{2\mu}p^2+\frac{1}{2}\mu\omega_c^2
q^2\right).
\end{equation}
Here $\Phi_0=\pi a^2_0 B_0$ is the flux of the magnetic field
${\bf B}^{(0)}_{in}$ inside the circle I which comes from the
second term of $p_1$ and $p_2$ in Eq.~(\ref{Eq:p-x-1}). That is,
$\Phi_0$ is only contributed by the AB vector potential ${\bf
A}^{(0)}_{out}$ in Eq.~(\ref{Eq:A}).
We introduce an annihilation operator
$A= \sqrt{\mu\omega_c/2\hbar}\;q +i\sqrt{1/2\hbar\mu\omega_c}\;p$
and its conjugate one $A^\dagger.$ The operators $A$ and
$A^\dagger$ satisfies $[A,A^\dagger]=1$. The eigenvalues of the
number operator $N=A^\dagger A$ is $n=0, 1, 2, \cdots$. Using $A$
and $A^\dagger$
to rewrite $J_z$, we obtain
%
%\begin{equation}
%\label{Eq:J-1}%e
%$J_z=\frac{q}{2\pi c}\Phi_0+\hbar\left(A^\dagger
%A+\frac{1}{2}\right).$
%
$J_z=q\Phi_0/2\pi c + \hbar\left(A^\dagger A+1/2\right).$
%\end{equation}
%
The zero-point angular momentum of $J_z$ is
$\mathcal{J}_0=\hbar/2+q\Phi_0/2\pi c.$ In the above the term
\begin{equation}
\label{Eq:J-AB}%13e
\mathcal{J}_{AB}=\frac{q}{2\pi c}\Phi_0
\end{equation}
is the fractional zero-point angular momentum of the ion
\cite{note-1} induced by the AB vector potential ${\bf
A}^{(0)}_{out}$. The magnetic flux $\Phi_0$ can be continuously
changed. This leads to $\mathcal{J}_{AB}$ taking fractional values
and the ground state of the angular momentum being infinitely
degenerate.

We notice that two vector potentials ${\bf A}^{(c)}_{in}$ and
${\bf A}^{(0)}_{out}$ play different roles: ${\bf A}^{(c)}_{in}$
contributes to energy spectra, but does not contribute to the
fractional angular momentum $\mathcal{J}_{AB}$; On the other hand,
${\bf A}^{(0)}_{out}$ does not contribute to energy spectra, but
contributes to $\mathcal{J}_{AB}$.

One point that should be emphasized is that $\mathcal{J}_{AB}$ is
only contributed by the the AB vector potential ${\bf
A}^{(0)}_{out}$ in Eq.~(\ref{Eq:A}). The structure of ${\bf
A}^{(0)}_{out}$ is special. Any other types of vector potentials
cannot contribute to $\mathcal{J}_{AB}$.

It can be proved that the fractional zero-point angular momentum
induced by the AB vector potential cannot be gauged away by a
gauge transformation \cite{JZZ08}. It is a real physical
observable.

{\bf Dynamics in the Intervening Region} -- In this region the
magnetic fields
${\bf B}^{(0)}_{out}={\bf B}^{(c)}_{out}=0,$
but there are two vector potentials of the AB type
$A^{(0)}_{out,i}$ and $A^{(c)}_{out,i}$.
The Hamiltonian of an ion in this region is
$\tilde H(x_1,x_2)=\left(\tilde K_1^2+\tilde K_2^2\right)/2\mu$
where $\tilde K_i$ is the mechanical momenta
\begin{equation}
\label{Eq:K2}%14e
\tilde K_i\equiv p_i+
\mu\omega_0 a^2_0\frac{\epsilon_{ij}x_j}{2\rho_0^2}+
\mu\omega_c a^2_c\frac{\epsilon_{ij}(x_j-x_{C,j})}{2\rho_c^2}.
\end{equation}
The Lagrangian corresponding to $\tilde H$ is
\begin{equation}
\label{Eq:L3}%15e
\tilde L=\frac{1}{2}\mu\dot{x_i}\dot{x_i}-\mu\omega_0
a^2_0\frac{\epsilon_{ij}\dot{x_i}x_j}{2\rho_0^2}
-\mu\omega_c
a^2_c\frac{\epsilon_{ij}\dot{x_i}(x_j-x_{C,j})}{2\rho_c^2}
\end{equation}
The $\tilde K_i$'s commute each other
\begin{equation}
\label{Eq:K-K}%16e
[\tilde K_i,\tilde K_j]=0.
\end{equation}
Behavior of $\tilde H$ is similar to a Hamiltonian of a free
particle. Its spectrum is a continuous one.

We emphasize again that vector potentials
${\bf A}^{(0)}_{out}$ and ${\bf A}^{(c)}_{out}$
of the AB type do {\it not} contributes to the commutators between
$\tilde K_i$'s. This does not lead to they contributing to energy
spectra of charged particles either.

We consider the limiting case of $\tilde H$ approaching to some
constant energy $\mathcal{\tilde E}_k$:
$\tilde H\to \tilde H_0=\mathcal{\tilde E}_k.$
The Lagrangian corresponding to $\tilde H_0$ is
\begin{equation}
\label{Eq:L4}%17e
\tilde L_0=-\mu\omega_0
a^2_0\frac{\epsilon_{ij}\dot{x_i}x_j}{2\rho_0^2}
-\mu\omega_c
a^2_c\frac{\epsilon_{ij}\dot{x_i}(x_j-x_{C,j})}{2\rho_c^2}
-\mathcal{\tilde E}_k.
\end{equation}
From $\tilde L_0$ it follows that the canonical momenta is
\begin{equation}
\label{Eq:p2}%18e
\tilde p_{i}=\frac{\partial\tilde {L_0}}{\partial \dot{x_i}}
=-\mu\omega_0 a^2_0\frac{\epsilon_{ij}x_j}{2\rho_0^2}
-\mu\omega_c a^2_c\frac{\epsilon_{ij}(x_j-x_{C,j})}{2\rho_c^2}.
\end{equation}
Eq.~(\ref{Eq:p2}) does not determine velocities $\dot{x_i}$ as
functions of $\tilde p_{i}$ and $x_j$, but gives the following
primary constraints
\begin{equation}
\label{Eq:C2}%19e
\tilde \varphi_i\equiv\tilde p_{i}+
\mu\omega_0 a^2_0\frac{\epsilon_{ij}x_j}{2\rho_0^2}
+\mu\omega_c a^2_c\frac{\epsilon_{ij}(x_j-x_{C,j})}{2\rho_c^2}=0.
\end{equation}
Here the special feature is that the corresponding Poisson
brackets between $\tilde \varphi_i$'s are identically zero,
\begin{equation}
\label{Eq:Poisson-2}%20e
\tilde C_{ij}=\{\tilde{\varphi_i}, \tilde{\varphi_j}\}\equiv 0.
\end{equation}
Because of $\{\hat \varphi_i, \hat \varphi_i\}$ identically
vanishing, it follows that the conditions of the constraints
$\tilde\varphi_i$ holding at all times lead to secondary
constraints
$\tilde\varphi_i^{(2)}=-\mu\omega_P^2 x_i$.
The Poisson brackets
$\{\tilde{\varphi_i}^{(2)}, \tilde{\varphi_j}\}=0$,
$\{\tilde{\varphi_i}^{(2)}, \tilde{\varphi_j}^{(2)}\}=0$,
and
$\{\tilde{\varphi_i}^{(2)}, \tilde H_0\}=0$,
so that persistence of the secondary constraints
$\tilde\varphi_i^{(2)}$ in course of time does not lead to further
secondary constraints $\tilde\varphi_i^{(3)}$.

Because of $\tilde C_{ij}\equiv 0,$ the inverse matrix
$\mathcal{\tilde C}^{-1}$ does not exist. The Dirac brackets
$\{\tilde{\varphi_i}, x_j\}_D$, $\{\tilde{\varphi_i}, p_j\}_D$,
$\{\tilde{\varphi_i}^{(2)}, x_j\}_D$, $\{\tilde{\varphi_i}^{(2)},
p_j\}_D$, $\{x_i, x_j\}_D$, $\{p_i, p_j\}_D$, and $\{x_i, p_j\}_D$
cannot be defined. According to Dirac's formalism of quantizing a
system with constraints, there is no way to establish dynamics at
the quantum mechanical level.

Properties of the AB vector potentials in the intervening region
is summarized as follows.
The intervening region is multiply connected. As is well known,
due to the non-trivial topology in this region, the interference
spectrum of charged particles suffered a shift according to the
quantum phase, i. e.
the amount of the loop integral of the AB vector potential around
an unshrinkable loop.
But unlike in the region II with a ``spectator" magnetic field, in
the intervening region the two AB vector potentials do not
contribute to physical observables:

(i) The AB vector potentials $A^{(0)}_{out,i}$ and
$A^{(c)}_{out,i}$ appear in the mechanical momenta $\tilde K_i$ of
Eq.~(\ref{Eq:K2}) and the corresponding Hamiltonian $\tilde H$,
but do not contribute to the commutators between $\tilde K_i$'s.
Therefore, they do not contribute to the energy spectrum. The
spectrum of the Hamiltonian $\tilde H$ is a continuous one of a
free particle.

(ii) In the limit of the Hamiltonian $\tilde H$ approaching some
constant the system has not a non-trivial dynamics survived at the
full quantum mechanical level.
Therefore, the two AB vector potentials ${\bf A}^{(0)}_{out}$ and
${\bf A}^{(c)}_{out}$ cannot contribute to the fractional angular
momentum $\mathcal{J}_{AB}$ by means of the constraints
$\tilde\varphi_i$'s of Eq.~(\ref{Eq:C2}).

{\bf Quantum Transmission of Information via Aharonov - Bohm
effect} -- Now we elucidate the
%physical meanings and an
application of the above results in the quantum transmission of
information.

The situation in the circle II is different from the intervening
region because of the ``spectator" magnetic field ${\bf
B}^{(c)}_{in}$.
The vector potentials ${\bf A}^{(c)}_{in}$ guarantees that in
Eq.~(\ref{Eq:Poisson-1}) the Poisson brackets of the constraints
$\varphi_i$ are well defined, which lead to a non-trivial dynamics
surviving at the full quantum mechanical level in the limit of the
Hamiltonian approaching its one of eigenvalues.
The vector potential ${\bf A}^{(0)}_{out}$ does not contribute to
energy spectra, but in this case it contributes to the fractional
angular momentum $\mathcal{J}_{AB}$ by means of the constraints
$\varphi_i$'s.
It is clear that though the vector potential ${\bf A}^{(c)}_{in}$,
like a ``spectator", does not contribute to $\mathcal{J}_{AB}$, it
plays essential role in guaranteeing non-trivial dynamics at the
quantum mechanical level.

If at a moment $t$ we adjust the magnetic field ${\bf
B}^{(0)}_{in}(t)$ in the circle I, though the magnetic field ${\bf
B}^{(0)}_{out}$ is zero outside the circle I everywhere.
According to the continuous condition of vector potentials, ${\bf
A}^{(0)}_{in}$ passes continuously over into ${\bf A}^{(0)}_{out}$
on the boundary between the circle I and the outer region.
At some later time $t+T$ a fractional angular momentum
$\mathcal{J}_{AB}(t+T)$ induced by the AB vector potential ${\bf
A}_{out}^{(0)}(t+T)$ in the circle II will be changed
correspondingly. Information encoded in variations of ${\bf
B}^{(0)}_{in}$ in the circle I is transmitted, moving through the
intervening region III, to $\mathcal{J}_{AB}$ in the circle II.

One point that should be emphasized is that this type of quantum
transmission has to satisfy the following condition: In the limit
of the Hamiltonian $H$ approaching one of its eigenvalues there is
non-trivial dynamics survived at the full quantum mechanical
level.
Here ``the full quantum mechanical level" means that in the
defined limit the reduced system can be quantized according to
Dirac's formalism of quantizing a system with constraints.
Information encoded in ${\bf A}^{(0)}_{out}$ moves through the
intervening region III, but cannot be received by an ion in this
region.
The reason is: because in the region III in the defined limit the
reduced system cannot be quantized according to Dirac's formalism
of quantizing a system with constraints. Thus there is no way to
establish dynamics at the full quantum mechanical level.
Therefore, the vector potential ${\bf A}^{(0)}_{out}$ cannot
contribute to the fractional angular momentum $\mathcal{J}_{AB}$
by means of the constraints $\tilde\varphi_i$'s of
Eq.~(\ref{Eq:C2}).
Such an intervening region may be called the blind area.

A special advantage of this type of transmission via the AB vector
potentials is that it is {\it noiseless}.
It is true that no quantum systems really isolated, and the
coupling to the uncontrollable environments produces noise.
Here the point is that the fractional angular momentum
$\mathcal{J}_{AB}$ is contributed {\it only} by the second term of
Eq.~(\ref{Eq:p-x-1}), which is related to the AB vector potential
$A^{(0)}_{out,i}$ of Eq.~(\ref{Eq:A}).
Generally, stray magnetic fields of environments very both in
space and time. Their vector potentials are not the AB type of
Eq.~(\ref{Eq:A}). Therefore, they cannot influence the
$\mathcal{J}_{AB}$.
Specially, the stray magnetic fields ${\bf B}^{(0)}_{s}$ in the
circle I and ${\bf B}^{(c)}_{s}$ in the II of environments add,
respectively, extra terms in Eq.~(\ref{Eq:p-x-1}) which are
related to the vector potentials $A^{(0)}_{s,i}$ and
$A^{(c)}_{s,i}$.
These terms contribute to the energy spectrum of the reduced
system in the required limit, but they do not influence the
$\mathcal{J}_{AB}$.
Another point that should be clarified is that there is {\it no}
energy transmission via the AB vector potential from the circle I
to the circle II.
Alternating electromagnetic fields induced by variations of ${\bf
B}^{(0)}_{in}$ in circle I transmit energy and interact with
environments. However, vector potentials of alternating
electromagnetic fields are not the AB type. They cannot influence
$\mathcal{J}_{AB}$ either.
The quantum transmission via vector potentials of the AB type is
not influenced by stray magnetic fields of the uncontrollable
environments and so on. It is {\it noiseless}.
The above investigation opens a way of quantum transmission of
information via the AB effect, leads to new experimental studies
and
%even potential
technological applications.
We expect results obtained in this paper to be of importance for
the current efforts in the field of quantum communication.

\vspace{0.4cm}

This work has been supported by the Natural Science Foundation of
China under the grant number 10575037 and by the Shanghai
Education Development Foundation.


\begin{thebibliography}{99}
\bibitem{note-1}
%H.J. Lipkin, M. Peshkin, Phys. Lett. {\bf 118B} (1982) 385.
%
%In this paper the fractional angular momentum of the
%electromagnetic fields was firstly revealed.
%%%%%%%%%%%%%%%%%%%%%%%%%%%%%%%%%%%%%%%%%%%%%%%%%%%%%%%%%%%%%%%%%%
Ref.~\cite{LP82} investigated the angular momentum ${\bf J}$
originated from the Poynting vector produced by crossing the
Coulomb field ${\bf E}$ of a charged particle with an external
magnetic field ${\bf B}$,
%%%%%%
$${\bf J}=
\frac{1}{4\pi c}\int{\bf r}\times[{\bf E}\times{\bf B(r)}]d^3 r.$$
%%%%%%
In cases where the magnetic field is only in the $z$-direction,
this angular momentum reduces to %
%%%%%
$$J_z=-\frac{q\phi}{2\pi c},$$
%%%%%%
where
%%%%%%
$\phi=\int\int B_z(x_1,x_2)dx_1dx_2$
%%%%%%
is the total magnetic flux. $J_z$ is the angular momentum of {\it
the electromagnetic fields}. In cases where the magnetic field
$B_z$ is produced by an infinitely long solenoid, this angular
momentum exists {\it only} inside the solenoid. $J_z$ should be
distinguished from $\mathcal{J}_{AB}$ of Eg.~(\ref{Eq:J-AB}).
$\mathcal{J}_{AB}$ is the angular momentum of {\it the ion} which
%It is worth noting that $\mathcal{J}_{AB}$
is induced by the AB vector potential
%on a charged particle
outside the solenoid. $\mathcal{J}_{AB}$ does not exist inside the
solenoid.
\end{thebibliography}
\end{document}